# High Resolution IRAS Maps and IR Emission of M31
## — I. Morphology and Sources

**Cong Xu**

Max-Planck-Institut für Kernphysik, Postfach 103980, 69117 Heidelberg, Germany

**George Helou**

IPAC 100-22, California Institute of Technology, Pasadena, CA 91125

July 23, 1995

Send correspondence to C. Xu

# ABSTRACT

The morphology of the infrared (IR) emission and the properties of discrete far-infrared (FIR) sources in M31 disk are studied using the high resolution (HiRes) maps from IRAS. Very thin and bright FIR arm segments are shown in these maps, which have similar structure as the HI gas but with much enhanced arm/inter-arm contrast, typically a factor of $\sim 5$ on the $60\mu m$ image. We identify 39 unconfused sources (excluding the nucleus) and 14 confused sources (7 pairs) at 60 $\mu$m by direct Gaussian fittings to the image. IRAS colors of 10 bright isolated sources are studied, which are presumably representative of the discrete sources in general. The colors follow the well known anticorrelation between $f_{60\mu}/f_{100\mu}$ and $f_{12\mu}/f_{25\mu}$. All sources coincide with optical HII regions. A comparison with H$\alpha$ observations shows that the total luminosity (in the wavelength range $8 - 1000\mu m$) associated with HII regions is $7.2 \pm 2.9 \; 10^8 L_\odot$, namely $30 \pm 14\%$ of the total IR emission of M31. Half of the IR luminosity of this component emerges from the sources. This luminosity translates into a present-day star formation rate of 0.36±0.14 $M_\odot$/yr, about an order of magnitude lower than that of the Milky Way Galaxy.

*Subject headings*: galaxies: individual – galaxies: interstellar matter – galaxies: photometry – interstellar: grains – interstellar: HII regions

# 1. Introduction

Two characteristics make M31, the Andromeda Galaxy, unique for a far-infrared (FIR) study. The first is its proximity: being the nearest spiral galaxy outside the Milky Way, it offers the best linear resolution for observations of spirals with a given angular resolution. This is a significant advantage for FIR observations, given the relatively poor angular resolution of the current generation of FIR instruments. Second, it is known to have a very low current star formation rate, perhaps an order of magnitude lower than that of the Milky Way (Walterbos 1988). Because of their very low FIR luminosities, not many such quiescent spirals can be studied in the FIR, and none except M31 with such high spatial resolution.

The FIR emission of M31 has been studied by Habing et al. (1984), Soifer et al. (1986), and very extensively by Walterbos and Schwering (1987, hereafter WS87) using IRAS maps. The main results from these early studies can be summarized as follows:

(1) M31 emits only a few percent of its total luminosity at wavelength longer than $10 \mu m$. The rest is mainly the optical radiation from stars (Habing et al. 1984; WS87).

(2) In the central part of the IRAS maps, corresponding to the optical bulge, the 12 and $25 \mu m$ radiation is due to circumstellar dust emission from late type stars, while the 60 and 100 $\mu m$ emission is due to the interstellar dust heated by the interstellar radiation field (ISRF) powered by the old bulge stars (Soifer et al. 1986).

(3) The disk IR emission is strongly concentrated in a ring with a semimajor axis of $\sim 45'$ and coincides in detail with the distribution of HII regions (Habing et al. 1984; see also Devereux et al. 1994). When it is decomposed into a warm (40K) component associated with star formation regions and a cool component associated with 'cirrus' (Cox et al. 1986; Helou 1986), less than 10 percent is due to the former (WS87).

(4) The strong mid-infrared (12 and 25 $\mu m$) excess and the rather constant 60–to–100 $\mu m$ ratio in the disk show evidence of the existence of very small grains/large molecules (WS87; Helou 1986; Xu & Helou 1994).

Recently, Devereux et al. (1994) made a comparison between the FIR emission and the H$\alpha$ emission of M31, and argued that the close correspondence between the FIR and the H$\alpha$ images suggests a common origin for the two emissions. According to these authors, up to 70% of the FIR emission of M31 could be associated with massive star formation regions, directly conflicting to the earlier result of WS87 (see above item 3).

We are carrying out a new FIR study on M31, utilizing the high resolution maps (of $\sim 1'$ resolution) produced with the HiRes software (Aumann, Fowler & Melnyk 1990; Rice 1993). Our motivation is to study the properties of FIR sources and the diffuse emission in the M31 disk *separately*. This was not possible in the earlier studies because of the relatively coarse resolution of IRAS maps. In an earlier paper (Xu & Helou 1994), we studied the IRAS color-color diagrams of a complete sample of small areas ($2' \times 2'$) in the M31 disk, and found evidence for a deficiency of Very Small Grains (but not the Polycyclic Aromatic Hydrocarbon molecules) in the areas dominated by the diffuse emission.

In this and a companion paper, we present a study on the general properties of FIR sources and of the diffuse emission in M31. In this paper, Paper I, we study the overall morphology of the FIR emission and the properties of the discrete sources in the M31 disk. In order to address the controversy as to the extent to which massive ionizing stars are responsible for the FIR emission of M31 (WS87; Devereux et al. 1994), we estimate the ratio between the FIR and H$\alpha$ fluxes of HII regions using those $60\mu m$ source which are fully covered in the H$\alpha$ survey of Walterbos and Braun (1992, hereafter WB92). We present the new HiRes maps of M31 in Section 2. The procedure of source extraction is described in Section 3. The FIR colors of some 'isolated' sources are studied in Section 4. In Section 5 we calculate the total dust emission of M31, from the sources and from the diffuse component respectively, and an estimate of the present-day star formation rate of M31 is made. Section 6 contains the summary. Throughout this paper, we assume for M31 a distance of 690 kpc ($1' = 200$ pc along the major axis), an inclination angle of $i = 77°$ ($i = 0$ for seen face-on), and P.A. $= 37°$.

## 2. HiRes IRAS maps

The high resolution IRAS maps of M31 at $12\mu m$, $25\mu m$, $60\mu m$ and $100\mu m$ are new IRAS products from the Infrared Processing and Analysis Center (IPAC), using the IPAC High-Resolution (HiRes) processor (Aumann, Fowler & Melnyk 1990; Rice 1993). After the 20th iteration of the deconvolution, the resolution achieved is $0'.9 \times 0'.5$ (in cross-scan and in-scan directions respectively) for the 12 and $25\mu m$ maps, $1'.0 \times 0'.8$ for the $60\mu m$ map, and $1'.5 \times 1'.5$ for the $100\mu m$ map (see also Rice 1993). However, the resolution is not uniform across the maps (Fowler & Aumann 1993), and depends in particular on the surface brightness of the background. In order to overcome this problem, and also to simplify the comparison between the four maps, we smooth all of them to a $1'.7$ circular beam on a grid with $0.5'$ pixels. The details of the production of the maps, and a comparison with previous M31 IRAS maps, can be found in Xu & Helou (1994; see also Xu & Helou 1995, hereafter ERRATUM).

These smoothed maps are presented in Figures 1 – 4. Allowing for the difference in sensitivity, the four maps have similar morphology, characterized by a bright nucleus-plus-ring structure with the ring coinciding with that of the HII regions (Pellet et al. 1978) and the nucleus coinciding with the optical bulge. This has been noticed in previous studies (Habing et al. 1984; WS87). The new HiRes maps however reveal features with a linear scale of a few hundred pc to $\sim 1$ kpc, namely the bright sources (giant star formation regions) and the very narrow widths of bright arms. Indeed it is rather surprising that the arms are so thin that they are hardly resolved by the HiRes beams in the cross-arm direction even on the original HiRes maps (resolution of $\sim 1'$). Compared to previous results (e.g. WS87), the contrast between the arm segments (including the 'ring') and the inter-arm region is significantly enhanced in the new HiRes maps (e.g. the $60\mu m$ map), with a typical value of about a factor of 5 (avoiding the bright sources). Unfortunately, one cannot exclude the possibility that the contrast is artificially exaggerated by HiRes. These results will however be tested in the upcoming ISO mission.

In the four panels of Fig.5, we compare the surface brightness distribution of the $60\mu m$ map and that of a $1.'7$ resolution HI map (Brinks 1984; Brinks & Shane 1984) along two cuts, running through the center of M31 at position angles of $24°$ and of $50°$, respectively. It appears that on the two maps there is good correspondence between the positions of features (dips and bumps), and the bumps representing the arm segments have about the same widths. On the other hand the HI emission decreases toward the center of the galaxy, while the $60\mu m$ emission does not. In comparison with optical maps of similar angular resolution (Walterbos & Kennicutt 1987), the FIR nucleus is smaller than the optical bulge, and the disk emission is much less smooth in the FIR than in the optical. An anti-correlation between the optical and the FIR surface brightness can be found in some places, e.g. the dust lanes which appear as bright thin arms on the FIR map, a natural consequence of the dust absorbing optical radiation and re-emitting in the FIR. Devereux et al. (1994) compared the HiRes FIR ($40 - 120\mu m$) map with the H$\alpha$ CCD image obtained with the Case Western Burrell Schmidt telescope at Kitt Peak, and found "striking correspondence" between the two.

## 3. Source extraction

There are many bright discrete sources on the HiRes maps. The brightest ones are concentrated on the famous 'ring' at about 9 kpc ($45'$) galactocentric radius. Compared to the HII regions map (Pellet et al. 1978; WB92; Devereux et al. 1994), these sources, except the nucleus, coincide exclusively with bright HII region complexes. A comparison between the $60\mu m$ sources and the optical HII regions (WB92) reveals that individual

FIR sources include typically 3 to 5 bright HII regions. The companion elliptical galaxy M32 does not show up in any of the four IRAS maps (Habing at al. 1984). Bright background radio sources (Berkhuijsen et al. 1983), which do not belong to the M31 disk, are not detected, either.

Because of the coarse resolution of IRAS, no analysis on discrete FIR sources in M31 could be carried out in previous studies. Even with the HiRes maps, many sources are still affected by confusion, in particular on the $100 \mu m$ map where they are more extended (Fig.4) and the spatial resolution is the poorest. The $12 \mu m$ and $25 \mu m$ maps have low signal-to-noise levels. Hence an overall source extraction is done for the $60 \mu m$ map only. We at first run an automatic source subtraction program which fits individual single elliptical Gaussians over areas of $4'.5 \times 4'.5$. For a source extraction threshold at a surface brightness of 1 MJy/sr above the background set by the local diffuse emission, we found 70 sources (excluding the nucleus) on the $60 \mu m$ map (Xu and Helou 1993). The limit was assigned because the one-$\sigma$ random fluctuation of the diffuse emission in the $60 \mu m$ disk of M31 is of the order of 0.5 MJy/sr. Therefore the sources are reliable at $\gtrsim 95$ percent confidence level ($\geq 2\sigma$ level). Because of the source confusion problem, the parameters (e.g. the major axis and minor axis, the flux, and even the detection) for many sources vary significantly when the size of the fitting area is changed. In order to assess the uncertainty due to this factor, we repeat the source extraction program using two more sizes of fitting area, i.e. $3' \times 3'$ and $6' \times 6'$. Out of the 70 sources detected in the first run (with the size of the fitting area of $4'.5 \times 4'.5$), 39 are detected in both of these two latter runs. They are represented by crosses in Fig.3.

For each of the 39 sources we have estimated, and reported in Table 1, the values and uncertainties of the following parameters: the coordinates R.A. (1950) and DEC (1950), the peak value of the $60 \mu m$ surface brightness $S_p$, the major (maj) and minor (min) axes, and the position angle of the major axis P.A. (measured from the North towards the East). For any parameter x, the value is estimated from the weighted mean of the logarithms of corresponding results in the 3 different fittings:

$$<\ln(x)> = \frac{\sum_{i=1}^{3} \omega_i \ln(x_i)}{\sum_{i=1}^{3} \omega_i} \qquad (1)$$

where $x_i$ represents the value of the parameter found in one of the fittings, and $\omega_i$ the weight:

$$\omega_i = \frac{1}{(\sigma_i/x_i)^2} \qquad (2)$$

where the $\sigma_i$ is the uncertainty on $x_i$. The variance of the logarithm is estimated with the same weight, and the systematic uncertainties due to the different Gaussian fittings are allowed for by introducing a scaling parameter s (Eadie et al. 1977):

$$s = \frac{1}{n-1} \sum_{i=1}^{n} \left[ \frac{\ln(x_i) - <\ln(x)>}{(\sigma_i/x_i)} \right]^2 = \frac{1}{n-1} \sum_{i=1}^{n} \omega_i \left[\ln(x_i) - <\ln(x)>\right]^2 , \qquad (3)$$

and

$$\sigma^2(\ln(x)) = s \times \frac{\sum_{i=1}^{n} \omega_i (\sigma_i/x_i)^2}{\sum_{i=1}^{n} \omega_i} = \frac{n}{n-1} \frac{\sum_{i=1}^{n} \omega_i \left[\ln(x_i) - <\ln(x)>\right]^2}{\sum_{i=1}^{n} \omega_i} , \qquad (4)$$

with n = 3. The values and uncertainties listed in Table 1 are calculated from these means and variances:

$$x = e^{<\ln(x)>} \qquad (5)$$

$$\sigma(x) = x\sqrt{\sigma^2(\ln(x))} . \qquad (6)$$

The uncertainties of the peak surface brightness and of the total flux of each source include also the calibration uncertainty of the $60\mu m$ map, which is at 4% level (Xu & Helou 1994).

The columns of Table 1 are arranged as follows:

Col.(1): source ID in the list;

Col.(2): Line 1: the 1950 equinox right ascension (R.A.) in hours, minutes and seconds;
          Line 2: uncertainty on R.A. in seconds of time;

Col.(3): Line 1: the 1950 equinox declination (DEC) in degrees, arcminutes and arcseconds;

Line 2: uncertainty on DEC in arcseconds;

Col.(4): Line 1: peak $60\mu m$ surface brightness $S_p$ in MJy/sr;

Line 2: uncertainty on $S_p$;

Col.(5): Line 1: major axis of the source in arcminutes;

Line 2: uncertainty on the major axis;

Col.(6): Line 1: minor axis of the source in arcminutes;

Line 2: uncertainty on the minor axis;

Col.(7): Line 1: position angle of the major axis in degree;

Line 2: uncertainty on the position angle;

Col.(8): Line 1: $60\mu m$ flux in Jy;

Line 2: uncertainty on the $60\mu m$ flux;

Col.(9): Line 1: sum of the H$\alpha$ fluxes of the corresponding HII regions in $10^{-12}\,\mathrm{erg/cm^2/s}$, taken from WB92;

Line 2: uncertainty of the H$\alpha$ flux;

Col.(10): ID's of the corresponding HII regions in the catalog of WB92 for ionized nebulae in the NE half of M31.

In addition, another 14 confused sources form 7 pairs (Table 2). For each pair two sources are detected individually in the test run with the fitting area size of $3'$ as well as in the original run with the fitting area size of $4'.5$, but in the run with the fitting size of 6' the two components are not distinguished, and therefore only one large source is detected. These confused sources are marked by open diamonds on the $60\mu m$ map (Fig.3) and listed in Table 2. The combined flux $f_{60\mu}$ and its uncertainty is calculated for each pair using the similar formulae as given in Eq's (1)–(6) with $f_{60\mu,1} = f^a_{60\mu,1} + f^b_{60\mu,1}$ and $f_{60\mu,2} = f^a_{60\mu,2} + f^b_{60\mu,2}$, where subscripts 1 and 2 denote the values found in the Gaussian fittings with areas of $3' \times 3'$ and $4'.5 \times 4'.5$ respectively, and superscripts a and b the values for the component a and b. The flux $f_{60\mu,3}$ is taken from that of the single source (including both components) found in Gaussian fitting to the $6' \times 6'$ area. The $f_{60\mu}$ of source pair 4a+4b has a large uncertainty (78%) due to the large dispersion

TABLE 2.

List of Confused 60$\mu m$ Sources (Source Pairs)

| (1) | (2) | (3) | (4) | (5) |
|---|---|---|---|---|
| ID | RA (1950) | DEC (1950) | $f_{60\mu}$ | unc |
| | h  m  s | °  ′  ″ | Jy | Jy |
| 1a | 00 38 13.0 | 40 20 15.2 | | |
| 1b | 00 38 21.4 | 40 21 47.6 | | |
| 1a+1b | | | 4.43 | 0.34 |
| 2a | 00 37 56.7 | 40 27 04.4 | | |
| 2b | 00 38 09.4 | 40 28 49.0 | | |
| 2a+2b | | | 2.89 | 1.32 |
| 3a | 00 38 40.8 | 40 33 42.8 | | |
| 3b | 00 38 50.1 | 40 35 10.0 | | |
| 3a+3b | | | 5.28 | 0.79 |
| 4a | 00 40 00.8 | 40 43 47.4 | | |
| 4b | 00 39 54.8 | 40 45 50.9 | | |
| 4a+4b | | | 2.76 | 2.15 |
| 5a | 00 39 22.0 | 41 07 09.9 | | |
| 5b | 00 39 09.4 | 41 08 16.2 | | |
| 5a+5b | | | 2.51 | 1.13 |
| 6a | 00 41 60.0 | 41 11 07.1 | | |
| 6b | 00 42 12.4 | 41 14 42.3 | | |
| 6a+6b | | | 9.27 | 1.92 |
| 7a | 00 40 16.9 | 41 20 53.6 | | |
| 7b | 00 40 24.9 | 41 21 57.5 | | |
| 7a+7b | | | 9.65 | 0.90 |

among the 3 fittings (the factor s Eq(3) is large). Because of the confusion problem, for individual components of each pair we list only the right ascension and declination which are calculated from the weighted mean of the corresponding values found in the two Gaussian fittings using areas of $3' \times 3'$ and $4'.5 \times 4'.5$ respectively.

## 4. IRAS colors of isolated sources

From Table 1 we selected 10 bright ($f_{60\mu} > 1$ Jy) 'isolated' sources (Table 3) around which there are no other sources within twice the Gaussian radius. They therefore do not suffer the confusion problem seriously. The IRAS colors of these isolated sources will be studied in this section, which are presumably representative for the discrete sources in general.

At the position of each isolated $60\mu m$ source, we did source extractions on the other 3 maps (i.e. the $12\mu m$ map, $25\ \mu m$ map, and the $100\mu m$ map) using single elliptical Gaussian fittings. For each source and each map the source extraction is repeated 3 times, with the size of the fitting area equal to $3' \times 3'$, $4'.5 \times 4'.5$ and $6' \times 6'$ respectively. All sources are detected in each of the three maps at least in two of the three runs with different fitting areas. In Table 3 we give their fluxes and the uncertainties in the IRAS bands, calculated using similar formulae to those given in Eq's (1)–(6). Some sources were detected only in two of the three Gaussian fittings (again with the size of the fitting area equal to $3' \times 3'$, $4'.5 \times 4'.5$ and $6' \times 6'$ respectively) at wavelengths other than $60\mu m$. In these cases, the corresponding flux and the uncertainty are calculated from the results in the two successful fittings, and are marked by a flag ":". The uncertainties include the calibration uncertainties of the corresponding maps, which are at 11%, 9%, 4% and 10% levels for the $12\mu m, 25\mu m, 60\mu m$ and $100\mu m$ maps respectively (Xu & Helou 1994).

The columns of Table 3 are arranged as follows:

Col.(1): source ID in Table 1;

Col.(2): Line 1: $12\mu m$ flux $f_{12\mu}$ in Jy;

Line 2: uncertainty of $f_{12\mu}$;

Col.(3): Line 1: $25\mu m$ flux $f_{25\mu}$ in Jy;

Line 2: uncertainty of $f_{25\mu}$;

Col.(4): Line 1: $60\mu m$ flux $f_{60\mu}$ in Jy;

Line 2: uncertainty of $f_{60\mu}$;

Col.(5): Line 1: $100\mu m$ flux $f_{100\mu}$ in Jy;

TABLE 3.

IRAS Fluxes of Isolated Sources

| (1) | (2) | (3) | (4) | (5) |
|---|---|---|---|---|
| ID | $f_{12\mu}$ unc | $f_{25\mu}$ unc | $f_{60\mu}$ unc | $f_{100\mu}$ unc |
| | Jy | Jy | Jy | Jy |
| 1 | 0.51 | 0.47 | 3.06 | 13.43 |
| | 0.14 | 0.07 | 0.18 | 1.62 |
| 2 | 0.11: | 0.21: | 2.08 | 4.08 |
| | 0.01: | 0.04: | 0.09 | 0.46 |
| 3 | 0.35 | 0.44: | 1.53 | 5.46 |
| | 0.04 | 0.07: | 0.13 | 0.70 |
| 6 | 0.21: | 0.36 | 1.57 | 4.43 |
| | 0.04: | 0.07 | 0.08 | 0.56 |
| 10 | 0.30 | 0.43 | 1.88 | 9.08 |
| | 0.04 | 0.07 | 0.16 | 1.48 |
| 12 | 1.12 | 1.13 | 5.46 | 23.86 |
| | 0.24 | 0.18 | 0.54 | 4.81 |
| 27 | 0.28 | 0.40: | 2.58 | 10.44 |
| | 0.08 | 0.04: | 0.22 | 1.55 |
| 31 | 0.30: | 0.82 | 6.52 | 18.71 |
| | 0.05: | 0.09 | 0.34 | 3.59 |
| 38 | 0.27 | 0.34 | 1.58 | 5.22 |
| | 0.04 | 0.05 | 0.15 | 0.98 |
| 39 | 0.32: | 0.49 | 2.81 | 10.84 |
| | 0.05: | 0.06 | 0.15 | 1.26 |

Line 2: uncertainty of $f_{100\mu}$.

Fig.6 is an IRAS color-color diagram ($\log(f_{12\mu}/f_{25\mu})$ vs. $\log(f_{60\mu}/f_{100\mu})$) of these isolated sources, which shows a clear anti-correlation between the colors. This anti-correlation has been found for galaxies (Helou 1986), for regions surrounding very massive stars (Boulanger et al. 1988), and for a complete sample of small areas within the M31 disk (Xu and Helou 1994). The trend found by WS87 that regions of high $60\mu m$ to $100\mu m$ intensity ratio correspond to relative depressions in the $12\mu m$ to $100\mu m$ intensity ratio may also be of the same nature (Helou, Ryter & Soifer 1991).

The FIR color ratio ($f_{60\mu}/f_{100\mu}$) of the sources varies in the range from $\sim 0.2$, which is slightly higher than the color ratio of the diffuse emission (0.17, see Xu & Helou 1994), to 0.5, with a mean of $0.30\pm0.04$. The means of other two IRAS colors are $<f_{12\mu}/f_{60\mu}> = 0.14 \pm 0.02$ and $<f_{25\mu}/f_{60\mu}> = 0.19 \pm 0.02$.

The mean $<f_{60\mu}/f_{100\mu}> = 0.30 \pm 0.04$ corresponds to a color temperature of $26\pm1$ K ($\nu^2$ emissivity law). Cox & Mezger (1988) and Rice et al. (1990) also found that for sources in the Galaxy and in M33 the average color temperature is about 30K, rather than 40K as assumed in WS87 and in other early papers (e.g. Cox et al. 1986). On the other hand, it appears that on average M31 sources are probably cooler than the M33 sources which have a mean $f_{60\mu}/f_{100\mu} = 0.40 \pm 0.11$ (Rice et al. 1990), although the difference is not significant because of the large error bars. This possible difference would be consistent with the optical studies of M31 HII regions which show: (1) the lack of bright regions and a possible excess of faint regions compared to other nearby spiral galaxies (Kennicutt et al. 1989; WB92), and (2) the giant HII regions in M31 being more diffuse than the ones in the Galaxy and in M33 (Kennicutt 1984). It might also be due to the deficiency of the Very Small Grains in M31 disk (Xu & Helou 1994).

## 5. Luminosity of the HII-region-associated FIR component and the present-day star formation rate

The FIR sources in M31 coincide generally with giant-HII or HII-complexes. This agrees fully with the results in the literature. Cox & Mezger (1988) and Pérault et al (1989) used the longitudinal profiles of Galactic emission in IRAS bands to separate the FIR sources and the diffuse emission in the Galaxy, and found that all the sources are associated with bright Giant Molecular Clouds/HII complexes, spiral arm segments and the Galactic center. Rice et al. (1990) detected 19 sources in the IRAS maps of M33 ($3' \times 5'$ resolution), all coincident with HII regions cataloged at optical and radio wavelengths. For the Large Magellanic Cloud, with IRAS maps of resolution of $\sim 3' \times 3'$, Xu et al. (1992) found that the probability of coincidence between bright FIR sources and bright HII regions is better than 90 percent. On the other hand, it is very likely that some HII regions which are faint and do not cluster around any giant HII regions will not appear as bright FIR sources, namely that part of the HII-region-associated FIR emission may not be represented by the FIR sources.

We try to estimate quantitatively how much the HII-region-associated component contributes to the total $60\mu m$ emission of M31, through a comparison between the $60\mu m$ sources and the HII regions detected in the CCD H$\alpha$ observations of WB92 for some regions in the north-east part of M31. The idea is that such a comparison will yield a mean of the $f_{60\mu}$/H$\alpha$ ratio for HII regions. Multiplying this ratio with the total H$\alpha$ flux of M31 HII regions, an estimate of the total $60\mu m$ flux of the HII-region-associated component can be made. Of the sources presented in Table 1, we find 12 cases which fall in the region covered by the H$\alpha$ observations. For these 12 we show the sum of the H$\alpha$ fluxes (uncorrected for internal extinction) of HII regions located within the region (defined by the major and minor axes and the P.A.) of the corresponding $60\mu m$ source. Since the resolution of the H$\alpha$ observations is much higher than that of our IRAS maps, an FIR source usually encloses many HII regions, of which we list only the brightest few. A weighted mean (weight=$1/\sigma^2$) of $< f_{60\mu}/f(H\alpha) >= 1.5(\pm 0.2)\ 10^{12}$ Jy/(erg cm$^{-2}$ s$^{-1}$) is found for these sources.

It should be pointed out that some of the diffuse H$\alpha$ emission which is not catalogued as extended structures in WB92, and thus is not included in the above calculation, so that the H$\alpha$ fluxes may be underestimated and the ratio overestimated. According to Walterbos and Braun (1994), the diffuse emission contributes 40% of the total H$\alpha$ emission of M31, some of which is included in the HII region catalogue as extended structures, and is not therefore missed by our estimate. On the other hand, the H$\alpha$ emission associated with the 60$\mu m$ sources is clearly dominated by a few very bright discrete HII regions. We therefore conclude that the error due to the omission of some diffuse H$\alpha$ emission cannot be very significant. There are eight more 60$\mu m$ sources in Table 1 in the north-east half of M31, whose IDs are flagged with stars. As noted in Table 1, they are either completely or partially outside the H$\alpha$ survey.

Walterbos and Braun (1994), extrapolating from their survey over the NE part of M31, estimate that a total H$\alpha$ luminosity of 1.2 $10^{40}$ erg s$^{-1}$ (uncorrected for internal extinction) is radiated from the star formation regions in M31. This includes the flux from the diffuse ionized gas (DIZ) found near the regions of star formation which is likely to be photo-ionized by massive stars (Walterbos and Braun 1994), but excludes the nucleus and the diffuse component in the inner part of the M31 disk which appears not associated with star formation regions (Devereux et al. 1994). Devereux et al. (1994) find a higher $H\alpha$ luminosity: excluding the nuclear emission and the diffuse emission in the inner disk, a total $H\alpha$ luminosity of 1.8$\pm$0.5 $10^{40}$ erg s$^{-1}$ (uncorrected for internal extinction) is observed from the 'star formation ring' in the M31 disk. However, this latter luminosity includes the emission from [NII] lines. Using the spectral data of eight Sab – Sbc galaxies in Kennicutt (1992), we find a mean H$\alpha$/(H$\alpha$ + [NII]) flux ratio of 0.61$\pm$0.13. Hence the result of Devereux et al. (1994) implies a total 'pure' H$\alpha$ luminosity of 1.1$\pm$3 $10^{40}$ erg s$^{-1}$ radiated from M31 star formation regions, consistent with the estimate of Walterbos and Braun (1994). In the following we will use this value as the estimate of the total H$\alpha$ flux from M31 star formation regions.

At the adopted distance of 690 kpc, the H$\alpha$ luminosity of 1.1 $10^{40}$ erg s$^{-1}$ corresponds to a flux of 1.9 $10^{-10}$ erg cm$^{-2}$ s$^{-1}$. Multiplying this H$\alpha$ flux with the mean $<f_{60\mu}/f(H\alpha)>$ ratio found for the 60$\mu m$ sources, we estimate that a 60$\mu m$ flux of 289 Jy, with a relative uncertainty of $\sim$ 40% (including both the error of the H$\alpha$ luminosity and that of the $<f_{60\mu}/f(H\alpha)>$ ratio), is due to dust associated with HII regions. This is 49$\pm$19 percent of the total 60$\mu m$ flux of M31, and twice the sum of f$_{60\mu m}$ of all sources listed in Table 1 and Table 2. Using this result and assuming that the mean IRAS colors of the isolated sources found in Section 4 represent those of the HII-region-associated component in general, we estimate the fluxes of the HII-region-associated dust in the other three IRAS bands which, together with $f_{60\mu}$, are presented in Table 4.

TABLE 4.

IR Fluxes and Luminosities

of HII-region-associated Component and of Diffuse Component

|  | $f_{12\mu}$ Jy | $f_{25\mu}$ Jy | $f_{60\mu}$ Jy | $f_{100\mu}$ Jy | fir[a] $10^{-8}$ erg cm$^{-2}$ s$^{-1}$ | $L_{IR}$[b] $10^9 \, L_\odot$ |
|---|---|---|---|---|---|---|
| HII | 37±15 | 52±21 | 289±115 | 974±390 | 2.17±0.87 | 0.72±0.29 |
| diffuse[c] | 131±51 | 79±42 | 306±118 | 2099±502 | 3.64±0.96 | 1.70±0.48 |
| total[d] | 168±49 | 131±36 | 595±28 | 3073±316 | 5.81±0.40 | 2.42±0.56 |

[a] Integrated FIR flux between $40 - 120\mu m$.

[b] Integrated IR luminosity between $8 - 1000\mu m$.

[c] Including the nucleus.

[d] IRAS fluxes are different from Xu & Helou (1994). See the erratum (Xu & Helou 1995).

From the IRAS fluxes listed in Table 4, and applying the formula of Helou et al. (1988), we find that the HII-region-associated component contributes 2.17(±0.87) $10^{-8}$ erg cm$^{-2}$ s$^{-1}$ of the integrated flux in the wavelength range 40–120$\mu m$, namely 37±15% of the total flux in this wavelength range radiated from M31 (Table 4). In order to estimate its contribution to the total dust emission ($8 - 1000 \, \mu m$) we need to extrapolate both the HII-region-associated component and the diffuse component to longer wavelengths beyond 120$\mu m$. Assuming that the very small grains are only half as abundant in M31 cirrus as they are in Galactic cirrus (Xu & Helou 1994) and applying a model based on that of Désert et al. (1990) with the ISRF intensity equal to that in the Solar Neighborhood (Fig.7), we find that the total dust emission for the diffuse component is $1.13(\pm 0.32)\,10^{-7}$ erg cm$^{-2}$ s$^{-1}$, 3.1 times that of its integrated flux in the wavelength range 40–120$\mu m$. The average spectrum of M31 sources is extended to the same wavelength range (Fig.7) by using another model based on that of Désert et al. for IR emission near an O5 star with the dilution factor $X_{O5} = 0.003$ (i.e. dust heated at 18 pc away from the star), and with the assumption that both the very small grains and the PAHs are only half as abundant in dust associated with M31 HII regions as they are in Galactic cirrus (PAHs may be destroyed in HII regions, Boulanger et al. 1988). We then find that the total dust emission of the HII-region-associated component is $4.8(\pm 1.9)\,10^{-8}$ erg/cm$^2$/sec, 2.2 times that of its integrated flux in the wavelength range 40–120$\mu m$. In terms of luminosity in the wavelength range of 8—1000$\mu m$, we conclude that M31 radiates $2.42(\pm 0.56)\,10^9 L_\odot$, of which $7.2(\pm 2.9)\,10^8 L_\odot$ (i.e. $30 \pm 14\%$) is due to the HII-region-associated component (Table 4).

It is interesting to note that the total IR emission of M31 found here is very close to the result of WS87 who found a total IR emission of $2.6\,10^9 L_\odot$. WS87 used a simpler two-component model with the warm component being a modified black body (assuming a $\nu^2$ dust emissivity law) specified by a single temperature of 40K, and the cool component a fixed spectrum with the $I_{60\mu}/I_{100\mu} = 0.16$ ($T_d = 21$K). On the other hand, the total IR luminosity of the HII-region-associated component is a factor 3.6 times the luminosity of the warm component of WS87, which is $2\,10^8 L_\odot$. The low luminosity of the warm component of WS87 is likely to be a consequence of the high color temperature (40K corresponding to $f_{60\mu}/f_{100\mu} \simeq 1$) adopted for the warm dust.

From the total luminosity of the HII-region-associated component, we can estimate the present-day star formation rate of M31 using the formula given in Walterbos (1988): $R_{IR} = 1 \times 10^{-3} L_{IR}(HII)/t_{IR}$, where $R_{IR}$ is the star formation rate in $M_\odot$/yr, $L_{IR}(HII)$ the total luminosity of the HII-region-associated component, and $t_{IR}$ the time scale over which massive stars contribute to the heating of HII-region-associated dust. Assuming $t_{IR} = 2 \times 10^6$ yr (Thronson & Telesco 1986), we find a present-day star formation rate of $R_{IR} = 0.36(\pm 0.14)\,M_\odot$/yr, well in the range of 0.2 – 0.5 $M_\odot$/yr found by Walterbos (1988) using several star formation indicators. This is about an order of magnitude lower than the present-day star formation rate of the Milky Way Galaxy (Mezger 1988).

Our results show that, despite such a low present-day star formation rate, a substantial part (~30—40%) of the FIR emission of M31 is due to the young high mass ionizing stars also responsible for the HII regions. For the Milky Way Galaxy, the contribution from the HII-region-associated component to the total FIR emission is only 20–30% (Cox & Mezger 1988; Pérault et al. 1989; Bloemen et al. 1990). The relatively high contribution from the HII-region-associated component to the FIR emission of M31 is probably due to a depression of the diffuse FIR emission of interstellar dust not associated with HII regions, which in turn is caused by 1) a deficiency of the non-ionizing UV radiation which is responsible for most of the heating of diffuse dust in an average spiral galaxy (Xu 1990), given that the 2000Å-to-blue flux ratio of M31 is one of the lowest in a large sample of nearby galaxies (Buat & Xu 1995); 2) a deficiency of diffuse dust in M31, corresponding to a face-on V-band optical depth of $\tau_V \sim 0.3$ (Paper II) which is more than a factor of 2 lower than the average face-on optical depth of Sb/Sc galaxies (Xu & Buat 1995). On the other hand, our results, showing that about 60% of the FIR ($40 - 120\mu m$) emission of M31 is due to diffuse dust not associated with HII regions, do not support the suggestion of Devereux et al. (1994) that the FIR emission of M31 is predominantly due to heating by ionizing stars.

## Summary

The high resolution (HiRes) IRAS maps of the Andromeda Galaxy (M31) are used to investigate the morphology of the infrared (IR) emission and the properties of discrete far-infrared (FIR) sources in its disk. Very thin and bright FIR arm segments are shown in these maps, which have similar structure as the HI gas but with much enhanced arm/inter-arm contrast, typically a factor of $\sim 5$ on the $60\mu m$ image. We identify 39 unconfused sources (excluding the nucleus) and 14 confused sources (7 pairs) at 60 $\mu$m by direct Gaussian fitting to the image, with careful treatment of the source-confusion problem. The IRAS colors of 10 bright isolated 60$\mu m$ sources are studied, as a presumably representative sample of the discrete source population. They follow the well known anticorrelation between $f_{60\mu}/f_{100\mu}$ and $f_{12\mu}/f_{25\mu}$, with mean IRAS colors $<f_{12\mu}/f_{60\mu}>= 0.14\pm 0.02$, $<f_{25\mu}/f_{60\mu}>= 0.19\pm 0.02$, and $<f_{60\mu}/f_{100\mu}>= 0.30\pm 0.04$. All sources except the nucleus coincide with optical HII regions. A comparison with H$\alpha$ observations shows that the total luminosity (in the wavelength range $8 - 1000\mu m$) associated with HII regions is $7.2\pm 2.9\ 10^8 L_\odot$, namely $30\pm 14\%$ of the total IR emission of M31. This is a factor of 3.6 greater than the IR luminosity of the warm component of Walterbos and Schwering (1987). However, our result does not agree with the suggestion of Devereux et al. (1994) that the FIR emission of M31 is predominantly energized by high mass stars. About half of the IR luminosity of the HII-region-associated component emerges from discrete FIR sources. The luminosity translates into a present-day star formation rate of 0.36$\pm$0.14 $M_\odot$/yr, about an order of magnitude lower than that of the Milky Way Galaxy.

*Acknowledgements.* We are very grateful to Dr. E. Brinks for providing the HI map of M31. We are indebted to the referee, Dr. R. Walterbos, whose comments helped to improve this paper in various aspects. Helpful discussions with C. Beichman, E. Berkhuijsen, and J. Fowler are acknowledged. Part of the work was done when CX was at Max-Planck-Institut für Radioastronomie, supported by an Alexander von Humboldt Fellowship. He thanks Prof. R. Wielebinski for his hospitality. This research is supported in part through the IRAS Extended Mission Program by the Jet Propulsion Laboratory, California Institute of Technology, under a contract with the National Aeronautics and Space Administration.

# References

Aumann, H. H., Fowler, J. W. & Melnyk, M., 1990, AJ 99, 1674

Beck, R., Gräve, R., 1982, A&A 105, 192

Berkhuijsen, E.M., Wielebinski, R., R. Beck, 1983, A&A 11, 141

Bloemen, J.B.G.M., Duel, E.R., Thaddeus, P., 1990 A&A 233, 437

Boulanger, F., Beichman, C., Désert, F.X., Helou, G. Pérault, M., Ryter, C. 1988, ApJ 332, 328

Brinks, E., 1984, Ph.D. thesis, University of Leiden

Brinks, E., Shane, W.W., 1984, A&AS 55, 179

Buat, V., Xu, C., 1995, A&A, in press

Cox, P., Mezger, P., 1988, in *Comets to Cosmology*, ed. A. Lawrence, Lecture Notes in Physics **297**, Springle, Heidelberg, p.97

Cox, P., Krügel, E., Mezger, P.G., 1986, A&A 155, 380

Désert, F.A., Boulanger, F., Puget, J.L., 1990, A&A 273, 215

Devereux, N.A. Price, R. Wells, L.A., Duric, N., 1994, AJ 108, 1667

Eadie, W.T., Drijard, D., James, F.E., Roos, M. Sadoulet, B. 1977, Statistical Methods in Experimental Physics (Amsterdam: North-Holland)

Fowler, J.H., Aumann, H.H., 1993, in *Science with High Spatial Resolution Far-Infrared Data*, eds. S. Terebey, J. Mazzarella, (Pasadena: Jet Propulsion Laboratory), p.1

Habing, H., et al., 1984, ApJ 278, L59

Helou, G., 1986, ApJ 311, L33

Helou, G., Khan, I.R., Malek, L., Beohmer, L., 1988, ApJS 68, 151

Helou, G., Ryter, C., Soifer, B.T., 1991, ApJ 376, 505

Kennicutt, R.C., 1984, ApJ 287, 116

Kennicutt, R.C., 1992, ApJ 388, 310

Kennicutt, R.C., Edgar, B.K., Hodge, P.W., 1989, ApJ 337, 761

Mezger, P.G., 1988, in *Galactic and Extragalactic Star Formation*, eds. R.E. Pudritz and M. Fich, Kluwer, Dordrecht, p.227

Pérault, M., Boulanger, F., Puget, J.L., Falgarone, E., 1989, unpublished

Pellet, A., Astier, N., Viale, A., Courtés, G., Maucherat, A., Monnet, G., Simien, F., 1978, A&AS 31, 439

Puget, J.L., Léger, A., 1989, ARA&A 27, 161

Rice, W., 1993, AJ 105, 67

Rice, W., Boulanger, F., Viallefond, F., Soifer, B.T., Freedman, W.L., 1990, ApJ 358, 418

Soifer, B.T., Rice, W.L., Mould, J.R., Gillett, F.C., Rowan-Robinson, M., Habing, H.J., 1986, ApJ 304, 651

Thronson, H.A., Telesco, C.M., 1986, ApJ 311, 98

Walterbos, R.A.M., 1988, in *Galactic and Extragalactic Star Formation*, eds. R.E. Pudritz and M. Fich, Kluwer, Dordrecht, p.361

Walterbos, R.A.M., Braun, R., 1992, A&AS 92, 625 (WB92)

Walterbos, R.A.M., Braun, R., 1994, ApJ 431, 156

Walterbos, R.A.M., Kennicutt, R.C., 1987, A&AS 69, 311

Walterbos, R.A.M., Schwering, P.B.W., 1987, A&A 180, 27 (WS87)

Xu, C., 1990, ApJ 365, L47

Xu, C., Buat, V., 1995, A&A 293, L65

Xu, C., Helou, G., 1993, in *Science with High Spatial Resolution Far-Infrared Data*, eds. S. Terebey, J. Mazzarella, (Pasadena: Jet Propulsion Laboratory), p.87

Xu, C., Helou, G., 1994, ApJ 426, 109

Xu, C., Helou, G., 1995, ApJ, in press (ERRATUM)

Xu, C., Klein, U., Meinert, D., Wielebinski, R., Haynes, R.F., 1992, A&A 257, 47

FIGURE CAPTIONS

**Figure 1.** Gray-scale map (overlaid with contours) of M31 at $12\mu m$. The beam size is $1'.7$ (the same in the following maps). The contour levels are 0.3, 0.9, 2.7 MJy/sr.

**Figure 2.** Gray-scale map (overlaid with contours) of M31 at $25\mu m$. The contour levels are 0.3, 0.9, 2.7 MJy/sr.

**Figure 3.** Gray-scale map (overlaid with contours) of M31 at $60\mu m$. The contour levels are 0.3, 0.9, 2.7, 8.1 MJy/sr. The unconfused sources (Table 1) are represented by the white crosses, each showing the location, size, and the orientation of a single source. The open diamond represent the confused sources (Table 2), each standing for a single component in a pair in Table 2.

**Figure 4.** Gray-scale map (overlaid with contours) of M31 at $100\mu m$. The contour levels are 1.2, 3.6, 10.8 MJy/sr.

**Figure 5.** Comparison between the $60\mu m$ surface brightness distribution (solid curve) and the HI column density distribution (dotted curve) along two cuts both passing through the center of M31. The HI column density is plotted with arbitrary units. Panel a: along the cut with the position angle P.A.=$50°$, in the NE half of M31. Panel b: along the same cut, in the SW half of M31. Panel c: along the cut with the position angle P.A.=$24°$, in the NE half of M31. Panel d: along the same cut, in the SW half of M31.

**Figure 6.** $R(60,100) = I_{60\mu}/I_{100\mu}$ v.s. $R(12,25) = I_{12\mu}/I_{25\mu}$ diagram of 10 isolated sources (see the text).

**Figure 7.** The IR spectra of M31 sources and of the diffuse component. The solid line is the spectrum calculated using a model based on that of Désert et al. (1990) for the IR emission near an O5 star with the dilution factor $X_{o5} = 0.003$, and assuming both Very Small Grains (VSGs) and PAHs are only half as abundant as they are in Galactic cirrus. The dashed line is to extrapolate the spectrum of the diffuse component to the whole IR wavelength range using a model based on that of Désert et al. (1990) for cirrus with the intensity of the IRSF equal to that in the Solar Neighborhood, and assuming the VSGs are only half as abundant as they are in Galactic cirrus.